\documentclass[9pt]{article}

\usepackage{geometry}
\usepackage{microtype}
\usepackage{subfigure}
\usepackage{booktabs} 

\usepackage{multirow}
\usepackage{graphicx}
\usepackage{amsmath,amsthm}
\theoremstyle{definition}

\usepackage{xspace,color}
\usepackage{diagbox,multirow} 
\usepackage{subfigure}
\usepackage{tikz,bm}
\usepackage{algorithmic,algorithm}
\definecolor{PineGreen}{rgb}{0.0, 0.47, 0.44}
\usepackage{enumitem}
\setlist[itemize]{leftmargin=*}
\usepackage{amsfonts}
\usepackage[utf8]{inputenc}
\usepackage[T1]{fontenc}    
\usepackage{url}            
\usepackage{booktabs}       
\usepackage{nicefrac}       
\usepackage{microtype}      
\usepackage{tikz,bm}
\usepackage{adjustbox}      
\usepackage{filecontents}
\usepackage{array,rotating}
\usepackage[labelfont=bf]{caption}

\usepackage{caption}
\usepackage{xspace,color}
\usepackage{enumerate}
\usepackage{enumitem,amsmath,amssymb}
\usepackage{hyperref}

\newcommand{\mbf}[1]{{\boldsymbol{\mathbf{#1}}}}
\renewcommand{\bm}{\mbf}

\newcommand{\mname}{\texttt{LINT}\xspace}

\usepackage{amssymb,bbm}
\usepackage{amsmath,amsthm,amsfonts,booktabs}
\usepackage[switch]{lineno}

\theoremstyle{definition}

\begin{document}
%
\title{Language Interaction Network for Clinical Trial Approval Estimation}

\author{Chufan Gao$^{1}$ \and Tianfan Fu$^{2}$ \and Jimeng Sun$^{1,3}$}
\date{%
    $^1$Department of Computer Science, University of Illinois Urbana-Champaign \\
    $^2$Department of Computer Science, Rensselaer Polytechnic Institute \\
    $^3$Carle Illinois College of Medicine, University of Illinois Urbana-Champaign
}

\maketitle

\begin{abstract}
Clinical trial outcome prediction seeks to estimate the likelihood that a clinical trial will successfully reach its intended endpoint. This process predominantly involves the development of machine learning models that utilize a variety of data sources such as descriptions of the clinical trials, characteristics of the drug molecules, and specific disease conditions being targeted. Accurate predictions of trial outcomes are crucial for optimizing trial planning and prioritizing investments in a drug portfolio. While previous research has largely concentrated on small-molecule drugs, there is a growing need to focus on biologics—a rapidly expanding category of therapeutic agents that often lack the well-defined molecular properties associated with traditional drugs. Additionally, applying conventional methods like graph neural networks to biologics data proves challenging due to their complex nature. To address these challenges, we introduce the Language Interaction Network (\mname), a novel approach that predicts trial outcomes using only the free-text descriptions of the trials. We have rigorously tested the effectiveness of \mname\ across three phases of clinical trials, where it achieved ROC-AUC scores of 0.770, 0.740, and 0.748 for phases I, II, and III, respectively, specifically concerning trials involving biologic interventions. 
\end{abstract}

\section{Introduction}
Accurate estimation of a clinical trial's success probability is essential for stakeholders such as researchers, biopharma investors, and others, informing their scientific and investment decisions. Inaccurate risk evaluation can lead to grave mistakes in drug development choices~\cite{wong2019estimation}. 
Moreover, given the high costs and generally low success rates of trials, it is crucial to prioritize correctly.
For example, approval rates for oncology drugs that enter clinical development are estimated to be as low as 3.4-19.4\%, 8.7-25.5\% for Cardiovascular, 8.2-15\% for Central Nervous System, etc~\cite{wong2019estimation, dimasi2010trends, lu2022cot,chen2021data}.

Drug research usually involves two phases: drug discovery and drug development. 
The goal of drug discovery is to design diverse and novel drug molecular structures with desirable pharmaceutical properties, while the goal of drug development is to evaluate the effectiveness and safety of the drug on human bodies via clinical trials. 

A drug needs to pass three phases of clinical trials to be approved and enter the medical market. 
Specifically, Phase I trials mainly focus on the safety and dosage of the drug molecules to human bodies (20 to 80 participants, several months, 70\% of drugs pass this phase), Phase II focuses on efficacy and side-effects (100 to 300 participants, could take several months to 2 years, 33\% of drugs pass this phase), and Phase III focuses on efficacy and monitoring of adverse reactions on broader population in treating disease (200 to 3,000 participants, 1 to 4 years, 25-30\% of drugs pass this phase)~\cite{fda_phase}. 


From a financial perspective, creating a novel drug-based treatment typically requires around 13–15 years and more than 2 billion dollars in research and development~\cite{martin2017much,fu2021differentiable,huang2022artificial,shen2023genocraft,chen2024congenital}. 
From 2009 to 2018, the FDA approved 355 new drugs and biologics.
As a concrete example to see the scale of the cost it takes to bring a drug to market: to treat heart failure, Novartis sponsored phase III of Entresto (a small-molecule drug), which recruited 4822 patients and spanned five years (2014-2019) which eventually yielded results that beneficial, but not statistically significant effects \footnote{\url{https://clinicaltrials.gov/ct2/show/study/NCT01920711}}~\cite{solomon2017angiotensin, solomon2019angiotensin,wang2024twin,lu2019integrated,fu2019ddl}. 
If machine learning models can predict the approval rate of a clinical trial before it starts, we could circumvent running high-risk trials that are likely to fail, which would save a significant amount of time and resources.  

Some of the main challenges for building accurate machine learning models for predicting trial outcomes are {\bf 1) limited training data and 2) diverse trial types}. 
Although clinical trial summaries are publicly available on \url{clinicaltrials.gov}, the limited number of labeled trials (success and failure) may be insufficient for training sophisticated machine learning models. 
Additionally, clinical trial descriptions vary significantly, with biologics often lacking typical molecular property information compared to conventional small-molecule drugs.

Despite these limitations, there are numerous opportunities for machine learning in this space. 
For example, a plethora of unstructured text is available that describes the various aspects of the trial and the drugs. 
Additionally, there also exists literature that describes the pharmaceutical properties of drug molecules (Absorption, Toxicity, etc). 

In recent years, there have been numerous breakthroughs in the natural language processing (NLP) community for clinical applications. 
For example, models can learn semantic knowledge from massive unlabeled data~\cite{lee2020biobert}, classify unstructured medical text with no human labels~\cite{gao2022classifying}, process clinical trial tabular data \cite{wang2022transtab, wang2023meditab}, and more \cite{wang2024reflection}. 
Inspired by recent trends in the NLP space, we propose a novel methodology: Language Interaction Network (\mname). 
Our model builds on pretrained language models (PLM) to predict trial outcomes by jointly considering the text descriptions of the trial, its associated drugs, and the corresponding medical codes.

We formally define \mname, a deep learning framework for clinical trial outcome prediction that can predict outcomes on \textit{both} small molecule drugs and biologics on Phase I, II, and III clinical trials taken from the largest labeled trial dataset.
It leverages text features and International Classification of Diseases (ICD) Codes to accurately predict the approval of interventional trials with \textit{both} small-molecule drugs and biologics. 
These text features include trial eligibility criteria, trial design specifications, drug property descriptions, and implicit PLM knowledge. 
In short, \mname learns a function on a weighted combination of pretrained large language model (LLM) embeddings to predict clinical trial outcomes. 

\begin{itemize}[leftmargin=*]
\item \textbf{Experimental results}. 
The proposed method achieves state-of-the-art performance and beats traditional models. 
Specifically, \mname\  obtains ROC-AUC scores of 0.723, 0.702, 0.770 (and F1 scores of 0.643, 0.654, and 0.740) on phase I, II, and III, respectively for clinical trials with biologics interventions, which are significantly better than other baselines (Section~\ref{sec:experiment}). 
\item \textbf{Open Source Code.} 
\mname\ is an open-source, flexible framework that is built on top of pretrained language models (PLM); thus, it is easily adapted when a novel PLM is released. 
The code will be publicly available at \url{https://anonymous.4open.science/r/LINT-E48D}. 

\item \textbf{Interpretability and Validation}. 
Although \mname\ is a deep neural network, we may use Shapley values (a measure of feature importance based on the average contribution of that word to the final prediction probability)
to interpret its decisions and visualize which portions of the input text are the most significant~\cite{fu2019pearl}. 
Furthermore, we will show in later sections that \mname's predicted score generally corresponds to the actual approval rate (see Section~\ref{sec:Results}).
\end{itemize}

\subsection{Related Work on Trial Outcome Prediction}
\label{sec:related_work}
Existing works often focus on predicting individual patient outcomes in a trial instead of a general prediction about the overall trial success. 
They usually leverage expert, hand-crafted features. 
Previous work has taken advantage of many aspects of the trial-related features \cite{Lo2019Machine, gayvert2016data, wu2012identifying, raj20evaluation, ye2009stochastic, siah2021predicting}.

Deep learning techniques have also seen widespread usage in learning representations from clinical trial data to support downstream tasks such as drug repurposing~\cite{huang2020deeppurpose,fu2021probabilistic,fu2020core}, patient retrieval~\cite{lu2023deep,gao2020compose} and enrollment~\cite{Biswal2020Doctor2VecDD,yi2018enhance}. Still, none have fully integrated the textual features along with the tabular trial information (such as allocation, primary purpose, and more) in a natural way.

Additionally, unlike previous work, the dataset used to train and test our proposed model is much larger than before, including both small molecule trials and biologic trials, extending the previous dataset used by Fu et al.~\cite{fu2022hint}. 
Also, these works generally optimize the representation learning for a single component in a trial, whereas \mname models a diverse set of trial text and tabular data jointly. Furthermore, previous methods will not work for biologics, which lack information that small-molecule drugs generally possess. 
\cite{lu2024uncertainty} enhance the accuracy of \cite{fu2022hint} by quantifying uncertainty of prediction.

\section{Methodology}
\label{sec:method}


\subsection{Formulation and Data Featurization}
In this section, we formulate the clinical trial outcome prediction problem into a binary classification problem. Additionally, we review all the data features, including trial eligibility criteria, drug descriptions, and disease codes.

Let the set of drug or biological interventions be denoted as: 
$$M = \{m_1, m_2, \dots, m_{N_M}\},$$
where $N_M$ is the total number of distinct interventions. 

Each drug $m_i=$ \{\textit{description, pharmacodynamics, toxicity, metabolism, absorption}\} also has plain text information regarding the ADMET properties and more (retrieved via DrugBank \cite{wishart2018drugbank})\footnote{\url{https://dev.drugbank.com/guides/fields/drugs}}: 

\begin{itemize}[leftmargin=*]
    \item \textit{description:} Brief summary about the drug--primary use cases, symptoms it treats;
    
    
    \item \textit{pharmacodynamics:} Description of how the drug works at a clinical or physiological level;	
    

    \item \textit{toxicity:} Lethal dose (LD50) values from test animals, description of side effects and toxic effects seen in humans;	
    \item \textit{metabolism:} Mechanism by which or organ location where the drug is neutralized; 	

    \item \textit{absorption:} Description of how much of the drug or how readily the drug is taken up by the body; 

    
\end{itemize}

Let the set of disease codes be denoted as 
$$D = \{d_1, d_2, \dots, d_{N_D}\},$$ 
where $N_D$ is the total number of distinct diseases, and each $d_i$ is an ICD code. 

Let each trial be represented as 
$$T = \{c_i, \dots, c_{n_c}, sum, d_i, \dots d_{n_d}, m_i, \dots, m_{n_m}, \bm{x}\},$$ 
where $c_i$ is the $ith$ inclusion/exclusion criterion sentence, $sum$ is the text summary of the trial, $d_i$ is the $ith$ associated ICD code, $m_i$ is the $ith$ associated drug/biologic intervention, and 

$\bm{x}$ = \{\textit{ec\_gender,
ec\_min\_age, ec\_max\_age, allocation, \\
intervention\_model, primary\_purpose, masking, sponsors, continent\}}

represents a feature vector of tabular features in the trial, described in the following section.

\textbf{Tabular Trial Data: }
Specifically, input is a list of text. First, each sentence $c_i$ are separated out in the trial criteria $\{c_i,\dots,c_{N_c}\}$ by splitting on new lines in the raw trial criteria text; these sentences are added to the list. Next, the brief summary information is appended to the list. Finally, additional tabular table information is processed and combined with the previous list (details in Appendix~\ref{app:trial_details}).

Note that each of the categorical variables also has an additional value for missing values. Finally, each of these categorical features is converted into text by the simple function template 
$\textbf{linearization}(\bm{x})$ = ``[feature\_name] [feature\_value]; \dots ; [feature\_name] [feature\_value]``.

\textbf{Data Processing for ICD Codes: }
To obtain a list of relevant ICD Codes, we use the API provided by \url{https://clinicaltables.nlm.nih.gov/} by inputting the conditions associated with each trial. 

\subsection{Neural Architecture}

\mname\  learns a classifier $\hat{y} = f_\theta (T),$
where $T$ is the trial with the associated drug intervention information as above. More specifically, \mname\  has 2 main modules, the transformer module, and the GRAM module (See Section~\ref{sec:Disease_code_representation}). 
$$f_\theta(T) = f_{\theta'}(CONCAT(h_{text}, h_{GRAM})),$$ 
where $f_{\theta'}$ is a classifier trained on the output of LINT's inner components, a transformer model that obtains embeddings 
$$h_{text}=f_{\theta^{text}}(c_i, \dots, c_{n_c}, sum, m_i, \dots, m_{n_m}, x)$$ and a GRAM model that obtains embeddings respectively 
$$h_{GRAM} = \frac{1}{n_d} \sum_{i=1}^{n_d} GRAM(d_i)$$ for a specific trial. Note that indexes for the arbitrary number of codes, eligibility criteria, and interventions are omitted.

\noindent\textbf{LINT Encoder: Text Data:}
This module handles all the text information in our input data and outputs embedding $h_{text}$.
We use a pretrained BERT model to extract the text features. BERT (Bidirectional Encoder Representations from Transformers) is a pretraining technique that captures language semantics and exhibits state-of-the-art performance in various NLP tasks~\cite{devlin2018bert}. Specifically, we use BioBERT~\cite{lee2020biobert} as implemented in the HuggingFace library~\cite{wolf-etal-2020-transformers}. BioBERT is pretrained on biomedical corpora--namely--PubMed abstracts and PubMed Central (PMC) full-text articles. This offers higher performance on domain-specific tasks such as biomedical relation extraction.

Due to the small number of trials, we use the transformer encoder to obtain a weighted mean over the original BioBERT embeddings. This allows \mname\  to take advantage of the powerful PLM while adding an additional layer of attention. 

Specifically, $f_{\theta^{text}}$ is consists of a PLM $LLM()$ and a transformer encoder $ENCODER()$. $LLM()$ simply takes in a varying-length text input and outputs a 768-dimensional embedding. $ENCODER()$ is a transformer encoder\footnote{\url{https://pytorch.org/docs/stable/generated/torch.nn.TransformerEncoder.html}} combined with a simple Linear layer that takes in a varying-length list of 768-dimensional inputs and outputs a scalar attention weight for all inputs, i.e., 
$$E = LLM(c_i, \dots, c_{n_c}, sum, m_i, \dots, m_{n_m}, x)$$ 
$$W = ENCODER(E)$$ 
$$f_{\theta^{text}} = E \cdot W,$$
where $E\in R^{B \times n}, W \in R^{n}$. $B$ is 768 in this case, and $n = n_c+1+n_m+len(x)$, i.e., the total length of the list of text inputs where $n_c$ is the number of eligibility criteria, $1$ represents the summary embedding, $n_m$ is the number of molecules, and $len(x)$ the length of the tabular data of the specific trial.

\noindent\textbf{GRAM Encoder: Disease Code Representation: }
\label{sec:Disease_code_representation}
This module handles all the ICD code information in our input data outputs embedding $h_{GRAM}$.
 
Disease codes are typically hierarchically organized in a directed tree. For example, ICD-10~\cite{10665-42980,lu2022cot} codes consist of three to seven characters. The beginning 3 characters represent the category to which the code belongs to: e.g., the range of A00-B99 consists of "Certain Infectious and Parasitic Diseases". The latter characters represent the specific condition: e.g., A15.0 represents "Tuberculosis of lung". 

We leverage graph attention-based model (GRAM)~\cite{choi2017gram} to represent disease code. 
Specifically, suppose $d$ is the disease code of interest, the ancestors of $d$ are $\mathcal{D}$. Then, GRAM yields an embedding for a given ICD code by the formulation: 
$ GRAM(d) = \sum_{d_j \in \mathcal{D} \cup \{d\}} \alpha_j d_j, $
where $\alpha_j$ are the learned attention weights for each level of ICD code.

Cross entropy loss is leveraged to guide the training 
$
L = \sum_{i=1}^{N} -y \log \hat{y}, 
$
where $y\in \{0,1\}$ and $\hat{y}\in (0,1)$ denote groundtruth and prediction, respectively. 

\textbf{\mname Classification:} Finally, the outputs of the 2 previous subsections are concatenated and input to a multilayer perceptron (MLP) for binary classification. Because our data is imbalanced, we use a weighted binary-cross-entropy loss, where the weight is calculated based on the label distribution in the training data.
$\ell(x, y) = -\sum_{c=1}^2 w_c \log \hat{y}_{c} y_{c}
$
Where $y_{i}$ is the predicted probability of class $i$ and $w_i$ is the class weight of $i$.

\section{Experiments}
\label{sec:experiment}

In this section, we present the experimental results. 
We start by describing the experimental setup, including dataset and splitting (Section~\ref{sec:data}). Finally, we present the trial outcome prediction results and analysis (Section~\ref{sec:Results}). 

\subsection{Data}
\label{sec:data}
All the historical clinical trial records can be downloaded from \url{https://clinicaltrials.gov/}. 
The trial success information is based on the benchmark from Fu et al. \cite{fu2022hint}\footnote{\url{https://github.com/futianfan/clinical-trial-outcome-prediction}}. 

\begin{table}[ht]
\caption{A breakdown of the trials after preprocessing of the 426,368 found on \url{clinicaltrials.gov}. No Label indicates the lack of success or failure label found. Non-Biological or Small-Molecule indicates that the drug does not have either of these interventions (i.e., those clinical trials may involve medical devices and behavioral interventions). All the labels are provided by Fu et al. \cite{fu2022hint}. A further breakdown of the labeled data is shown in Table~\ref{tab:data_splits}.}
\label{tab:Trial_breakdown}
\centering
\resizebox{0.5\linewidth}{!}{
    \begin{tabular}{c|c}
        Flag & Number of trials \\
        \hline
        No Label & 149,091 \\
        Non-Biological or Small-Molecule & 155,353 \\
        Non Interventional & 96,538 \\
        Labeled Trials & 25,386 \\
        \hline
    \end{tabular}
    }

\end{table}

\begin{table}[ht]
\centering
\caption{Table showcasing data partitions according to modality and phase. The final two columns provide the total quantity of training data for training and testing, and the percentage of successful trials—in parentheses—of all trials. Bio. refers to biologics, Drugs to small-molecule drug candidates, and Both to a combined set of biologic and small-molecule trials.}
\label{tab:data_splits}
\resizebox{0.5\linewidth}{!}{%
\begin{tabular}{c|c|cc}
Mode & Phase & \# Train (Pos. \%) & \# Test (Pos. \%) \\ \hline
\multirow{3}{*}{Bio.} & 1 & 505 (71.29\%) & 441 (71.88\%) \\
 & 2 & 973 (59.61\%) & 571 (55.52\%) \\
 & 3 & 692 (78.03\%) & 366 (74.32\%) \\ \hline
\multirow{3}{*}{Drugs} & 1 & 2032 (59.45\%) & 1681 (59.79\%) \\
 & 2 & 6401 (48.68\%) & 3873 (54.40\%) \\
 & 3 & 4745 (65.99\%) & 2388 (67.63\%) \\ \hline
\multirow{3}{*}{Both} & 1 & 2418 (62.57\%) & 2019 (62.95\%) \\
 & 2 & 6999 (50.49\%) & 4215 (55.26\%) \\
 & 3 & 5249 (67.96\%) & 2619 (69.19\%) \\ \hline
\end{tabular}
}
\vspace{-1em}
\end{table}

\noindent{\bf Trial data preprocessing:} Data preprocessing pared down the original total of 426,368 clinical trials to 23,519 valid trials for consideration. We focused on interventional trials, excluding observational ones, and further narrowed the scope to biological or drug interventions. Trials with efficacy concerns or lacking outcome labels were also omitted. The final breakdown comprised of 
(4437 Phase I, 11214 Phase II, and 7868 Phase III trials). 
See Table~\ref{tab:Trial_breakdown} for a detailed account of the number of trials post-preprocessing.

\begin{table*}[ht]
\centering
\caption{Phase 3 baseline comparisons. Combined indicates the combined set of Biologics and Drugs. }
\label{tab:bio_baselines}
\resizebox{.8\linewidth}{!}{%
\begin{tabular}{c|c|cccc}
Mode & Model & PR-AUC & ROC-AUC & F1 & Acc \\ \hline
\multirow{6}{*}{Biological} & Logistic Regression & 0.846 $\pm$ 0.029 & 0.697 $\pm$ 0.043 & 0.859 $\pm$ 0.015 & 0.777 $\pm$ 0.022 \\
 & SVM & 0.867 $\pm$ 0.019 & 0.692 $\pm$ 0.035 & 0.872 $\pm$ 0.014 & 0.785 $\pm$ 0.022 \\
 & Decision Tree & 0.797 $\pm$ 0.019 & 0.622 $\pm$ 0.028 & 0.799 $\pm$ 0.016 & 0.704 $\pm$ 0.022 \\
 & AdaBoost & 0.857 $\pm$ 0.022 & 0.705 $\pm$ 0.036 & 0.855 $\pm$ 0.010 & 0.767 $\pm$ 0.014 \\
 & Random Forest & 0.863 $\pm$ 0.016 & 0.731 $\pm$ 0.022 & 0.844 $\pm$ 0.012 & 0.755 $\pm$ 0.017 \\
 & LINT & \textbf{0.882 $\pm$ 0.016} & \textbf{0.770 $\pm$ 0.028} & \textbf{0.879 $\pm$ 0.010} & \textbf{0.817 $\pm$ 0.016} \\ \hline
\multirow{6}{*}{Drug} & Logistic Regression & 0.831 $\pm$ 0.011 & 0.703 $\pm$ 0.010 & 0.781 $\pm$ 0.010 & 0.691 $\pm$ 0.011 \\
 & SVM & 0.811 $\pm$ 0.011 & 0.681 $\pm$ 0.012 & 0.797 $\pm$ 0.009 & 0.692 $\pm$ 0.012 \\
 & Decision Tree & 0.717 $\pm$ 0.010 & 0.573 $\pm$ 0.010 & 0.717 $\pm$ 0.008 & 0.621 $\pm$ 0.009 \\
 & AdaBoost & 0.814 $\pm$ 0.012 & 0.692 $\pm$ 0.013 & 0.794 $\pm$ 0.007 & 0.700 $\pm$ 0.009 \\
 & Random Forest & 0.760 $\pm$ 0.013 & 0.635 $\pm$ 0.011 & 0.756 $\pm$ 0.009 & 0.658 $\pm$ 0.010 \\
 & HINT & $0.7338 \pm 0.0090$ & $0.6916 \pm 0.0144$ & $0.7922 \pm 0.0070$ & $0.6955 \pm 0.0089$ \\
 & LINT & \textbf{0.854 $\pm$ 0.010} & \textbf{0.740 $\pm$ 0.011} & \textbf{0.820 $\pm$ 0.008} & \textbf{0.726 $\pm$ 0.011} \\ \hline
\multirow{6}{*}{Combined} & Logistic Regression & 0.856 $\pm$ 0.009 & 0.732 $\pm$ 0.010 & 0.806 $\pm$ 0.007 & 0.717 $\pm$ 0.008 \\
 & SVM & 0.833 $\pm$ 0.007 & 0.699 $\pm$ 0.012 & 0.808 $\pm$ 0.005 & 0.705 $\pm$ 0.007 \\
 & Decision Tree & 0.724 $\pm$ 0.010 & 0.563 $\pm$ 0.013 & 0.720 $\pm$ 0.009 & 0.619 $\pm$ 0.011 \\
 & AdaBoost & 0.832 $\pm$ 0.009 & 0.704 $\pm$ 0.010 & 0.800 $\pm$ 0.007 & 0.705 $\pm$ 0.009 \\
 & Random Forest & 0.781 $\pm$ 0.011 & 0.647 $\pm$ 0.013 & 0.769 $\pm$ 0.006 & 0.670 $\pm$ 0.007 \\
 & LINT & \textbf{0.860 $\pm$ 0.009} & \textbf{0.748 $\pm$ 0.009} & \textbf{0.826 $\pm$ 0.005} & \textbf{0.737 $\pm$ 0.007}
\end{tabular}
}
\end{table*}
\begin{table*}[ht]
\centering
\caption{Results of \mname on the different data splits on the test set.} 
\label{tab:results}
\resizebox{.7\linewidth}{!}{%
\begin{tabular}{c|c|cccc}
Mode                       & Phase & PR-AUC       & ROC-AUC      & F1           & Acc.         \\ \hline

\multirow{3}{*}{Biologics}            & 1 & 0.860 $\pm$ 0.026 & 0.723 $\pm$ 0.029 & 0.778 $\pm$ 0.015 & 0.694 $\pm$ 0.018 \\
                                      & 2 & 0.758 $\pm$ 0.022 & 0.702 $\pm$ 0.011 & 0.687 $\pm$ 0.016 & 0.651 $\pm$ 0.011 \\
                                      & 3 & 0.882 $\pm$ 0.016 & 0.770 $\pm$ 0.028 & 0.879 $\pm$ 0.010 & 0.817 $\pm$ 0.016 \\
                                      \hline
\multirow{3}{*}{Small-Molecule Drugs} & 1 & 0.728 $\pm$ 0.014 & 0.643 $\pm$ 0.014 & 0.698 $\pm$ 0.008 & 0.615 $\pm$ 0.009 \\
                                      & 2 & 0.696 $\pm$ 0.010 & 0.654 $\pm$ 0.007 & 0.678 $\pm$ 0.008 & 0.606 $\pm$ 0.008 \\
                                      & 3 & 0.854 $\pm$ 0.010 & 0.740 $\pm$ 0.011 & 0.820 $\pm$ 0.008 & 0.726 $\pm$ 0.011 \\
                                      \hline
\multirow{3}{*}{Combined}             & 1 & 0.770 $\pm$ 0.015 & 0.667 $\pm$ 0.013 & 0.716 $\pm$ 0.010 & 0.637 $\pm$ 0.010 \\
                                      & 2 & 0.699 $\pm$ 0.010 & 0.650 $\pm$ 0.006 & 0.706 $\pm$ 0.006 & 0.585 $\pm$ 0.007 \\
                                      & 3 & 0.860 $\pm$ 0.009 & 0.748 $\pm$ 0.009 & 0.826 $\pm$ 0.005 & 0.737 $\pm$ 0.007
\end{tabular}
}

\end{table*}

\noindent{\bf Prediction by trial phase:} Predictions are reported by trial phase, reflecting the distinct objectives of each phase. 
Phase I primarily determines the safe maximum dosage, focusing on potential adverse effects. 
Phase II examines the efficacy, using metrics tailored to the intervention's aim - such as tumor size, survival rate, or quality of life.
This phase typically experiences the most noise, given the complex task of defining success. 
Phase III compares the new drug with existing treatments, often employing double-blind methodologies involving multiple treatment arms.

\noindent{\bf Train/Test Split:} To split the data into train, test, and validation sets, we manually chose the year 2015 as the cutoff date (specifically, January 1, 2015), following the methodology of~\cite{fu2022hint,chen2024uncertainty,fu2021probabilistic}.
This split strategy is to circumvent the information leakage because later trials are typically relying on the knowledge from the earlier trials. 
All trials whose completion date is confirmed to be before the cutoff are considered as training and validation data; otherwise, it is considered test data.

Finally, we also split by Phase and Intervention Type. 
We consider predicting trial outcomes from phases 1, 2, and 3. There are 3 interventions that we consider: only small molecule drugs, only biologics, or both. 
Table~\ref{tab:data_splits} shows the full information regarding splits based on drug type and phase.

Information regarding baselines--Logistic Regression, SVM, Decision Tree, AdaBoost, Random Forest, and HINT--are shown in Appendix~\ref{app:baseline_methods}. 

\subsection{Results \& Analysis}
\label{sec:Results}

In this section, we present and analyze the experimental results. 
Table~\ref{tab:results} reveals that \mname\ excels at Phase 3 prediction, garnering Test AUC scores of 0.770, 0.740, and 0.745 for biologic, drugs, and combined predictions, respectively. While the model delivers a solid performance in Phase 1 prediction, its overall performance dips in Phase 2. This downturn is expected, given that Phase 2 has the highest trial volume and is generally the most challenging task.

Table~\ref{tab:bio_baselines} comparison reveals \mname\ surpasses all baseline models in all metrics. The simple Logistic Regression with our BERT embedding input, the second-best model, achieves Phase 3 Test F1 scores of 0.865, 0.800, and 0.806, which still lags considerably behind \mname. Note that HINT is solely applicable in Drugs mode, as it doesn't cater to Biologics.

\noindent\textbf{Performance Breakdown by Disease Type:}
\begin{table*}[ht]
\centering
\caption{A table of popular ICD categories (taken from the trial's ICD code) for small molecule drugs, biologics, and both combined. 
N represents the number of samples, and (Hist. App. \%) represents the historical approval rate--the positive labels indicating trial successes. 
All metrics are from \mname.
(Note that the category ``Factors influencing health status and contact with health services'' includes codes that represent HIV infection status, family and personal history of cancer, family and personal history of diabetes, and cystic fibrosis carrier)}
\label{tab:Example_Diseases}
\resizebox{\linewidth}{!}{%
\begin{tabular}{c|c|ccccc}
\hline 
Mode & Category & N (Hist. App. \%) & Test PR-AUC & Test ROC-AUC & Test F1 & Test Acc. \\ \hline
\multirow{5}{*}{Biologics} & Neoplasms & 80 (37.50) & 0.603 $\pm$ 0.090 & 0.722 $\pm$ 0.048 & 0.435 $\pm$ 0.075 & 0.713 $\pm$ 0.042 \\
 & Certain infections and parasitic diseases & 64 (84.38) & 0.915 $\pm$ 0.049 & 0.674 $\pm$ 0.123 & 0.913 $\pm$ 0.032 & 0.848 $\pm$ 0.050 \\
 & \begin{tabular}[c]{@{}c@{}}Factors influencing health status and \\ contact with health services\end{tabular} & 47 (72.34) & 0.897 $\pm$ 0.061 & 0.870 $\pm$ 0.064 & 0.899 $\pm$ 0.039 & 0.855 $\pm$ 0.054 \\
 & Diseases of the respiratory system & 42 (83.33) & 0.943 $\pm$ 0.025 & 0.761 $\pm$ 0.084 & 0.926 $\pm$ 0.029 & 0.877 $\pm$ 0.045 \\
 & \begin{tabular}[c]{@{}c@{}}Diseases of the musculoskeletal \\ system and connective tissue\end{tabular} & 41 (78.05) & 0.846 $\pm$ 0.077 & 0.603 $\pm$ 0.114 & 0.880 $\pm$ 0.046 & 0.801 $\pm$ 0.068 \\ \hline
\multirow{5}{*}{\begin{tabular}[c]{@{}c@{}}Small-\\ Molecule \\ Drugs\end{tabular}} & Neoplasms & 568 (53.52) & 0.719 $\pm$ 0.029 & 0.709 $\pm$ 0.027 & 0.703 $\pm$ 0.023 & 0.670 $\pm$ 0.022 \\
 & \begin{tabular}[c]{@{}c@{}}Factors influencing health status and \\ contact with health services\end{tabular} & 373 (70.51) & 0.900 $\pm$ 0.019 & 0.798 $\pm$ 0.024 & 0.847 $\pm$ 0.014 & 0.765 $\pm$ 0.017 \\
 & \begin{tabular}[c]{@{}c@{}}Endocrine, nutritional and metabolic \\ diseases\end{tabular} & 345 (79.42) & 0.934 $\pm$ 0.022 & 0.787 $\pm$ 0.044 & 0.893 $\pm$ 0.016 & 0.814 $\pm$ 0.025 \\
 & Diseases of the nervous system & 322 (62.11) & 0.816 $\pm$ 0.022 & 0.723 $\pm$ 0.027 & 0.767 $\pm$ 0.016 & 0.648 $\pm$ 0.020 \\
 & Certain infections and parasitic diseases & 313 (73.80) & 0.877 $\pm$ 0.032 & 0.751 $\pm$ 0.033 & 0.873 $\pm$ 0.019 & 0.787 $\pm$ 0.029 \\ \hline
\multirow{5}{*}{Combined} & Neoplasms & 585 (53.68) & 0.752 $\pm$ 0.028 & 0.724 $\pm$ 0.023 & 0.672 $\pm$ 0.024 & 0.660 $\pm$ 0.018 \\
 & \begin{tabular}[c]{@{}c@{}}Factors influencing health status \\ and contact with health services\end{tabular} & 405 (71.60) & 0.902 $\pm$ 0.014 & 0.793 $\pm$ 0.019 & 0.847 $\pm$ 0.010 & 0.768 $\pm$ 0.013 \\
 & Certain infections and parasitic diseases & 360 (75.83) & 0.871 $\pm$ 0.021 & 0.730 $\pm$ 0.029 & 0.878 $\pm$ 0.018 & 0.792 $\pm$ 0.028 \\
 & \begin{tabular}[c]{@{}c@{}}Endocrine, nutritional and metabolic \\ diseases\end{tabular} & 351 (79.77) & 0.930 $\pm$ 0.017 & 0.779 $\pm$ 0.031 & 0.893 $\pm$ 0.012 & 0.816 $\pm$ 0.019 \\
 & Diseases of the nervous system & 334 (62.87) & 0.808 $\pm$ 0.027 & 0.709 $\pm$ 0.026 & 0.770 $\pm$ 0.019 & 0.659 $\pm$ 0.021
 \\ 
\hline 
\end{tabular}
}
\vspace{-1em}
\end{table*}
Table~\ref{tab:Example_Diseases} shows the top 5 most common ICD categories for each of the three modes (by occurrence in the clinical trial data). 
Note that the most common categories are naturally different for each modality. 
However, we can see that \mname\  generally performs well and achieves high accuracy and ROC-AUC over most categories. 
It is interesting to see that for Biologics interventions for Neoplasms (cancer-related trials), the F1 score is much lower than the accuracy. 
This indicates that our model tends to predict true negative samples better than true positives in the Neoplasm case. 
However, in most other categories, both and the accuracy is high.

Features are masked out by random sampling from the existing dataset when not considered for traditional tabular feature classification. 
In this case, masking is done by replacing words with the [MASK] token by the SHAP package. 
Section~\ref{sec:Case_Studies} shows an example of this. 
From this, we see that the model generally does rely on informative portions of text. 

\section{Conclusion}
Clinical trial outcome prediction is vital for predicting the safety of new drugs and biologics. 
In this paper, we focus on developing a machine learning model to predict the outcome of clinical trial that can account for biologics, a quickly growing intervention type. 
Specifically, we propose an open-source, flexible framework that is built on top of pretrained language models--\mname--a method that supports the accurate prediction of success in clinical trials.

Thorough empirical studies are carried out to validate the effectiveness of the proposed method, which achieves state-of-the-art ROC-AUC scores on predicting approval of phase III trials, beating many traditional and recent baselines. 
We validate the effectiveness of \mname\  with thorough experiments across three trial phases. 
Specifically, \mname\  obtains 0.770, 0.740, and 0.748 ROC-AUC scores on phase I, II, and III, respectively for clinical trials with biologic interventions. 
We also show that \mname is generally well calibrated and demonstrated \mname's performance on the top-5 most popular categories of ICD codes. 
Additionally, using Shapley values, we visualize the portions of the input text that are the most significant for the prediction of success/failure.

\textbf{Future Work:} Future research should address the substantial lack of clear outcome labels in clinical trial datasets, an issue currently unaddressed, with potentially over 100,000+ unlabeled trials, as indicated in Table~\ref{tab:Trial_breakdown}. Strategies employing unsupervised learning, such as masked language modeling, weak supervision, or semi-supervision could be instrumental in resolving this. It's also crucial to improve label quality, a challenging task due to the complex language in result descriptions. Endeavors to identify human-interpretable automatic labels could help expand the dataset.
Interpretability is also key for decision-makers. Understanding why a trial is predicted to fail or succeed, rather than simply knowing the outcome, is valuable. While \mname\ can be paired with Shapley values to highlight text sections affecting prediction confidence, it doesn't directly account for human-interpretability. Future research should focus on creating interpretable models, which could optimize clinical trial design and enhance success rates.

\section{Appendix}
\label{sec:appendix}

\subsection{Literature Review Extended}
In this section, we briefly review the related literature. 
In recent years, there have been several attempts to use machine learning to predict clinical trial outcomes.
\subsection{Traditional Machine Learning Methods}
Lo et al.~\cite{Lo2019Machine} applied traditional machine-learning techniques (penalized logistic regression (PLR), random forests (RF), neural networks (NN), gradient boosting trees (GBT), support vector machines (SVM)~\cite{scikit-learn}, and decision trees C.50~\cite{c50}) to predict drug approvals using drug and clinical trial data. 
Payvert et al.~\cite{gayvert2016data} introduced PrOCTOR, a random forest-based model to predict drug toxicity using 10 molecular properties, 34 target-based properties, and 4 drug-likeness rule features. 
Hong et al.~\cite{hong20predicting} designed an ensemble classifier based on weighted least squares support vector regression (LS-SVR) to predict the success/failure of clinical trials. 

Wu et al.~\cite{wu2012identifying} developed a two-stage SVM classification method to identify genes and genetic lesion statuses in clinical trials.
Raj et al.~\cite{raj20evaluation} used Gradient-Boosted Decision Trees (GBDT~\cite{ye2009stochastic}) to predict patients who responded to treatment on various depressive symptoms utilizing pretreatment symptom scores and electroencephalographic features.
Siah et al.~\cite{siah2021predicting} created an open challenge and compared over 3000 various machine learning models for clinical trial outcome prediction. 
They found that the best-performing model was an ensemble consisting of two XGBoost models~\cite{chen2015xgboost} and one Bayesian logistic regression (BLR) model~\cite{blr}.

Our proposed model \mname differs from the previously mentioned methods in several ways. 
Most do not consider drug molecule features and trial protocol information jointly, rather generally relying on sets of hand-annotated features. 
Furthermore, most trials suffer from small training data sources, whereas \mname takes into account a large, multi-modal dataset of text and tabular data. 

\subsection{Deep Representation Learning Related to Clinical Trials}

Recently, deep learning has been rising in popularity in the machine learning for healthcare space; specifically, it has been used to learn representation from clinical trial data to support downstream tasks such as drug repurposing~\cite{huang2020deeppurpose,fu2021probabilistic}, patient retrieval~\cite{zhang2020deepenroll,gao2020compose} and enrollment~\cite{Biswal2020Doctor2VecDD}. 

Doctor2Vec~\cite{Biswal2020Doctor2VecDD}, a recently proposed hierarchical clinical trial embedding where the unstructured trial descriptions were embedded using Bidirectional Encoder Representations from Transformers (BERT)~\cite{devlin2018bert}. 

DeepEnroll~\cite{zhang2020deepenroll} leverages a hierarchical embedding model to represent patient longitudinal electronic health record (EHR) and aligns it with eligibility criteria (EC) via a numerical information embedding and entailment module to reason over numerical information in both EC and EHR.

Gao et al.~\cite{gao2020compose} proposed a patient-trial matching model to find qualified patients for clinical trials given structured EHR and unstructured EC text with both inclusion and exclusion criteria. 
The core of this model consists of a convolutional highway network and a hierarchical memory network that generates a contextualized word embedding for each word of the trial protocol. 
Multiple one-dimensional convolutional layers with varying kernel sizes capture semantics at different granularity.


Qi et al.~\cite{pmlr-v106-qi19a} designed a Residual Semi-Recurrent Neural Network and took phase 2 results as features to predict the phase 3 outcome.  
This network consists of an RNN with a residual connection from the first input and performs significantly better than RNNs. 
The trough concentration (Ctrough) and Phase 2 subject--level baseline characteristics were used to build an individual treatment effect (ITE) model for Phase 3 trial patients.


Fu et al.~\cite{fu2022hint} designed a Hierarchical Interaction Network (HINT) to capture the interaction between multi-modal features (drug molecules, disease codes, eligibility). 
It uses an interaction graph module on embeddings produced via domain knowledge to capture various relations between EC, molecule structure, trial protocol, and more to predict trial outcomes. 
However, this work does not support biologics-related interventions because the lack of protein structures and molecule properties such as absorption, distribution, metabolism, excretion, and toxicity (ADMET) are not known.

\subsection{Baseline Methods}
\label{app:baseline_methods}
We employ the following baseline methods for clinical trial outcome prediction and compare them with our proposed \mname\ method. Each baseline method is upgraded with the same input embedding used in \mname, i.e., BERT embeddings of text and GRAM embeddings of the ICD codes for diseases addressed in the clinical trials.
\begin{itemize} 

\item \textbf{Logistic Regression} is a common model that models the log-odds for a class through a linear combination of the input features, similar to a simple one-layer neural network with a logistic activation function~\cite{Lo2019Machine}.

\item \textbf{Supporting Vector Machine (SVM)} is another common linear model that attempts to fit a maximum-margin hyperplane between the input features (often using a nonlinear kernel function) in order to separate classes 
\cite{wu2012identifying}.
\item \textbf{Decision Tree} is a hierarchical, rule-based model that's generally trained using algorithm that attempts to iteratively split on a feature using an information-theoretic measure like label entropy at each branch of the tree~\cite{Lo2019Machine}. 

\item \textbf{AdaBoost} is a meta-estimator that iteratively fits a decision tree and then fits additional copies of the classifier on the re-weighted dataset (where weights of incorrectly classified instances are increased to emphasize them)~\cite{fu2022hint}. 

\item \textbf{Random Forest} is an ensemble of decision trees trained on different sub-samples of the input data (usually via sampling with bootstrapping)~\cite{Lo2019Machine}.

\item \textbf{Hierarchical Interaction Network (HINT)} is the previous state-of-the-art model we compare against. It is a complex model, consisting of a graph attention network, highway networks, and more to combine drug structure, eligibility criteria, and ICD codes in order to make a binary trial success prediction~\cite{fu2022hint}. 
\end{itemize}

\subsection{Trial Details}
\label{app:trial_details}
We describe the processing for each of the tabular features in $\bm{x}$.
\begin{itemize}[leftmargin=*]
    \item \textit{ec\_gender:} The eligible patient genders (male/female / or either) that the trial considers from eligibility criteria.
    \item \textit{ec\_min\_age}, \textit{ec\_max\_age:} The minimum and maximum ages of patients selected via the eligibility criteria. Notes that we convert the valid age range of the eligibility criteria to 4 bins following the Research Inclusion Statistics Report from the NIH~\cite{national2021nih}. Ages below 6 are considered children, ages 6-18 are considered adolescents, and ages 18-65 are considered adults. Finally, ages higher than 65 are considered older adults. I.e., 4 bins of (<6, 6-18, 18-65 and >65).
    \item \textit{allocation:} the treatment allocation, which can be randomized or nonrandomized. 
    \item \textit{intervention\_model:} The general design of the strategy for assigning therapies and can be Crossover, Factorial, Parallel, Sequential, or Single Group Assignment.
    \item \textit{primary\_purpose:} Describes the trial purpose, including Basic Science, Diagnostic, Educational/Counseling/Training, Health Services Research, Prevention, Screening, Supportive Care, Treatment, or Other.
    \item \textit{masking:} The type of method (single, double, triple, or quadruple masking) used to keep the study group assignment hidden after allocation between parties (Participants, Care Providers, Investigators, and Outcomes Assessors).
    \item \textit{sponsors:} The organization that oversees the trial. Since there are hundreds of possible sponsors, we simply denote separate sponsors into Large or Small, where large sponsors are the top 10 most common sponsors over all trials \footnote{Top 10 sponsors: GlaxoSmithKline, Merck Sharp \& Dohme LLC, Sanofi Pasteur, a Sanofi Company, Amgen, Pfizer, National Cancer Institute (NCI), Novartis Pharmaceuticals, Abbott, Bristol-Myers Squibb, Novartis Vaccines}.
    \item \textit{continents:} The continents in which the study was performed were converted from the raw trial "country" data. 
\end{itemize}

\subsection{Ablations}
\begin{table*}[ht]
\centering
\caption{Ablation experiments on removing different parts of the text from the input data. The first row denotes the complete input data. The rest of the rows indicates results from removal of that specific text feature only (preserves the other inputs text features).}
\label{tab:ablations}
\resizebox{.7\linewidth}{!}{
\begin{tabular}{c|cccc}
\hline
Ablation          & PR AUC            & ROC AUC           & F1                & Acc.              \\ \hline
All Data Included & 0.766 $\pm$ 0.007 & 0.679 $\pm$ 0.005 & 0.726 $\pm$ 0.005 & 0.647 $\pm$ 0.005 \\
No Trial Summary  & 0.703 $\pm$ 0.011 & 0.613 $\pm$ 0.007 & 0.000 $\pm$ 0.000 & 0.386 $\pm$ 0.007 \\
No Trial Tabular Data & 0.770 $\pm$ 0.008 & 0.684 $\pm$ 0.006 & 0.632 $\pm$ 0.005 & 0.616 $\pm$ 0.004 \\
No EC             & 0.770 $\pm$ 0.008 & 0.683 $\pm$ 0.005 & 0.660 $\pm$ 0.005 & 0.625 $\pm$ 0.004 \\
No Drugs          & 0.771 $\pm$ 0.008 & 0.685 $\pm$ 0.006 & 0.635 $\pm$ 0.005 & 0.618 $\pm$ 0.004 \\ \hline
\end{tabular}
}
\end{table*}

We conduct ablations by excluding text features from the trial summary, trial tabular data, eligibility criteria, and drug information, as shown in Table~\ref{tab:ablations}. Interestingly, removing certain text features marginally boosts ROC-AUC performance, but at the expense of F1 and Accuracy. Furthermore, excluding the trial summary significantly impairs \mname's performance, with all metrics dropping. This could elucidate why removing other features doesn't impact the results, as the model appears to predominantly rely on the trial summary.

Despite this, \mname might still find other outcomes important, yet it primarily hinges on the trial summary, thereby heavily influenced if it's absent. Future studies should delve into this dependency since other text features also significantly contribute to trial outcome prediction.


\subsubsection{Case Studies}
\label{sec:Case_Studies}
In this section, we take a closer look at 2 different biologic drugs. 
Secukinumab and Botulinum toxin type A, two random examples chosen from the test set, for the fair evaluation of \mname.

\textbf{Case 1: Secukinumab (NCT02404350)}\footnote{\url{https://clinicaltrials.gov/ct2/show/study/NCT02404350}}~\cite{mease2018secukinumab,lu2022cot}. 
First, let us take a look at a successful test prediction. 
As a brief summary, Novartis Pharmaceuticals sponsored this study, and the main goal of this study was to demonstrate efficacy on inhibition of progression of structural damage of secukinumab in subjects with active Psoriatic Arthritis (PsA) as measured by improvement in physical function measured by Health Assessment Questionnaire and skin and nail improvement for psoriasis signs. 
This study was deemed successful and had P-value of less than 0.0001 in its Estimation Parameter of Odds Ratio. 
\mname \ predicted success with a normalized score of 0.79 (from 0 to 1).

\textbf{Case 2: Botulinum toxin type A (NCT02660359)}\footnote{\url{https://clinicaltrials.gov/ct2/show/results/NCT02660359}}
Here, we take a look at an incorrect test prediction. 
This study was sponsored by Ipsen, and the primary purpose of this trial was to analyze safety and efficacy of two Dysport~\cite{ranoux2002respective} (Similar to Botox) doses (600 units U and 800 U), compared to placebo in reducing urinary incontinence (UI) in adult subjects treated for neurogenic detrusor overactivity (NDO) due to spinal cord injury (SCI) or multiple sclerosis (MS). 
This study was annotated to be unsuccessful due to the lack of participants; however, upon further analysis of the trial, all p-values were significantly less than 0.05, indicating statistical significance. 
This could indicate that if given more participants, a successful outcome may have been possible. 
\mname \ predicted success with a normalized score of 0.38 (from 0 to 1).

\small
\bibliographystyle{plain}
\bibliography{custom}

\begin{thebibliography}{10}

\bibitem{Biswal2020Doctor2VecDD}
S.~Biswal et~al.
\newblock Doctor2vec: Dynamic doctor representation learning for clinical trial recruitment.
\newblock In {\em AAAI}, 2020.

\bibitem{chen2024congenital}
Jintai Chen, Shuai Huang, Ying Zhang, Qing Chang, Yixiao Zhang, Dantong Li, Jia Qiu, Lianting Hu, Xiaoting Peng, Yunmei Du, et~al.
\newblock Congenital heart disease detection by pediatric electrocardiogram based deep learning integrated with human concepts.
\newblock {\em Nature Communications}, 15(1):976, 2024.

\bibitem{chen2021data}
Lulu Chen, Chiung-Ting Wu, Robert Clarke, Guoqiang Yu, Jennifer~E Van~Eyk, David~M Herrington, and Yue Wang.
\newblock Data-driven detection of subtype-specific differentially expressed genes.
\newblock {\em Scientific reports}, 11(1):332, 2021.

\bibitem{chen2015xgboost}
Tianqi Chen, Tong He, Michael Benesty, Vadim Khotilovich, Yuan Tang, Hyunsu Cho, Kailong Chen, Rory Mitchell, Ignacio Cano, Tianyi Zhou, et~al.
\newblock Xgboost: extreme gradient boosting.
\newblock {\em R package version 0.4-2}, 1(4):1--4, 2015.

\bibitem{chen2024uncertainty}
Tianyi Chen, Nan Hao, and Capucine Van~Rechem.
\newblock Uncertainty quantification on clinical trial outcome prediction.
\newblock {\em arXiv preprint arXiv:2401.03482}, 2024.

\bibitem{choi2017gram}
Edward Choi et~al.
\newblock Gram: graph-based attention model for healthcare representation learning.
\newblock In {\em KDD}, 2017.

\bibitem{fda_phase}
Office of~the Commissioner.
\newblock Step 3: Clinical research.

\bibitem{devlin2018bert}
Jacob Devlin et~al.
\newblock {BERT}: Pre-training of deep bidirectional transformers for language understanding.
\newblock In {\em NAACL-HLT}, 2019.

\bibitem{dimasi2010trends}
Joseph~A DiMasi, Lanna Feldman, Abraham Seckler, and Andrew Wilson.
\newblock Trends in risks associated with new drug development: success rates for investigational drugs.
\newblock {\em Clinical Pharmacology \& Therapeutics}, 87(3):272--277, 2010.

\bibitem{fu2021differentiable}
Tianfan Fu, Wenhao Gao, Cao Xiao, Jacob Yasonik, Connor~W Coley, and Jimeng Sun.
\newblock Differentiable scaffolding tree for molecular optimization.
\newblock {\em ICLR}, 2022.

\bibitem{fu2022hint}
Tianfan Fu, Kexin Huang, Cao Xiao, Lucas~M Glass, and Jimeng Sun.
\newblock {HINT}: Hierarchical interaction network for clinical-trial-outcome predictions.
\newblock {\em Patterns}, 3(4):100445, 2022.

\bibitem{fu2021probabilistic}
Tianfan Fu, Cao Xiao, Cheng Qian, Lucas~M Glass, and Jimeng Sun.
\newblock Probabilistic and dynamic molecule-disease interaction modeling for drug discovery.
\newblock In {\em Proceedings of the 27th ACM SIGKDD conference on knowledge discovery \& data mining}, pages 404--414, 2021.

\bibitem{fu2020core}
Tianfan Fu, Cao Xiao, and Jimeng Sun.
\newblock Core: Automatic molecule optimization using copy \& refine strategy.
\newblock In {\em AAAI}, 2020.

\bibitem{gao2022classifying}
Chufan Gao, Mononito Goswami, Jieshi Chen, and Artur Dubrawski.
\newblock Classifying unstructured clinical notes via automatic weak supervision.
\newblock {\em arXiv preprint arXiv:2206.12088}, 2022.

\bibitem{gao2020compose}
Junyi Gao, Cao Xiao, Lucas~M Glass, and Jimeng Sun.
\newblock Compose: Cross-modal pseudo-siamese network for patient trial matching.
\newblock In {\em KDD}, 2020.

\bibitem{fu2019pearl}
Tian Gao, Cao Xiao, Tengfei Ma, and Jimeng Sun.
\newblock Pearl: Prototype learning via rule learning.
\newblock In {\em Proceedings of the 10th ACM International Conference on Bioinformatics, Computational Biology and Health Informatics}, pages 223--232, 2019.

\bibitem{gayvert2016data}
Kaitlyn~M Gayvert, Neel~S Madhukar, and Olivier Elemento.
\newblock A data-driven approach to predicting successes and failures of clinical trials.
\newblock {\em Cell chemical biology}, 23(10):1294--1301, 2016.

\bibitem{blr}
Ben Goodrich, Jonah Gabry, Imad Ali, and Sam Brilleman.
\newblock rstanarm: {Bayesian} applied regression modeling via {Stan}., 2020.
\newblock R package version 2.21.1.

\bibitem{fu2019ddl}
Trong~Nghia Hoang, Cao Xiao, and Jimeng Sun.
\newblock {DDL}: Deep dictionary learning for predictive phenotyping.
\newblock In {\em IJCAI: proceedings of the conference}, volume 2019, page 5857. NIH Public Access, 2019.

\bibitem{hong20predicting}
Zhen~Yu Hong, Jooyong Shim, Woo~Chan Son, and Changha Hwang.
\newblock Predicting successes and failures of clinical trials with an ensemble ls-svr.
\newblock {\em medRxiv}, 2020.

\bibitem{huang2020deeppurpose}
Kexin Huang et~al.
\newblock Deeppurpose: a deep learning library for drug--target interaction prediction.
\newblock {\em Bioinformatics}, 2020.

\bibitem{huang2022artificial}
Kexin Huang, Tianfan Fu, Wenhao Gao, Yue Zhao, Yusuf Roohani, Jure Leskovec, Connor~W Coley, Cao Xiao, Jimeng Sun, and Marinka Zitnik.
\newblock Artificial intelligence foundation for therapeutic science.
\newblock {\em Nature Chemical Biology}, 18(10):1033--1036, 2022.

\bibitem{c50}
Max Kuhn and Ross Quinlan.
\newblock {\em C50: C5.0 Decision Trees and Rule-Based Models}, 2023.
\newblock R package version 0.1.8.

\bibitem{lee2020biobert}
Jinhyuk Lee, Wonjin Yoon, Sungdong Kim, Donghyeon Kim, Sunkyu Kim, Chan~Ho So, and Jaewoo Kang.
\newblock Bio{BERT}: a pre-trained biomedical language representation model for biomedical text mining.
\newblock {\em Bioinformatics}, 36(4):1234--1240, 2020.

\bibitem{Lo2019Machine}
Andrew~W. Lo, Kien~Wei Siah, and Chi~Heem Wong.
\newblock Machine learning with statistical imputation for predicting drug approvals.
\newblock {\em Harvard Data Science Review}, 1(1), 7 2019.

\bibitem{lu2019integrated}
Yingzhou Lu, Yi-Tan Chang, Eric~P Hoffman, Guoqiang Yu, David~M Herrington, Robert Clarke, Chiung-Ting Wu, Lulu Chen, and Yue Wang.
\newblock Integrated identification of disease specific pathways using multi-omics data.
\newblock {\em bioRxiv}, page 666065, 2019.

\bibitem{lu2024uncertainty}
Yingzhou Lu, Tianyi Chen, Nan Hao, Capucine~Van Rechem, Jintai Chen, and Tianfan Fu.
\newblock Uncertainty quantification and interpretability for clinical trial approval prediction.
\newblock {\em Health Data Science}, 2024.

\bibitem{martin2017much}
Linda Martin et~al.
\newblock How much do clinical trials cost?
\newblock {\em Nat. Rev. Drug Discov.}, 16(6):381--382, June 2017.

\bibitem{mease2018secukinumab}
Philip Mease, D{\'e}sir{\'e}e van~der Heijde, Robert Landew{\'e}, Shephard Mpofu, Proton Rahman, Hasan Tahir, Atul Singhal, Elke Boettcher, Sandra Navarra, Karin Meiser, et~al.
\newblock Secukinumab improves active psoriatic arthritis symptoms and inhibits radiographic progression: primary results from the randomised, double-blind, phase iii future 5 study.
\newblock {\em Annals of the rheumatic diseases}, 77(6):890--897, 2018.

\bibitem{national2021nih}
National~Institutes of~Health et~al.
\newblock Nih rcdc inclusion statistics report, 2021.

\bibitem{10665-42980}
World~Health Organization.
\newblock Icd-10 : international statistical classification of diseases and related health problems : tenth revision, 2004.

\bibitem{scikit-learn}
F.~Pedregosa, G.~Varoquaux, A.~Gramfort, V.~Michel, B.~Thirion, O.~Grisel, M.~Blondel, P.~Prettenhofer, R.~Weiss, V.~Dubourg, J.~Vanderplas, A.~Passos, D.~Cournapeau, M.~Brucher, M.~Perrot, and E.~Duchesnay.
\newblock Scikit-learn: Machine learning in {P}ython.
\newblock {\em Journal of Machine Learning Research}, 12:2825--2830, 2011.

\bibitem{pmlr-v106-qi19a}
Youran Qi and Qi~Tang.
\newblock Predicting phase 3 clinical trial results by modeling phase 2 clinical trial subject level data using deep learning.
\newblock volume 106 of {\em Proceedings of Machine Learning Research}, pages 288--303, 2019.

\bibitem{raj20evaluation}
Pranav Rajpurkar et~al.
\newblock {Evaluation of a Machine Learning Model Based on Pretreatment Symptoms and Electroencephalographic Features to Predict Outcomes of Antidepressant Treatment in Adults With Depression: A Prespecified Secondary Analysis of a Randomized Clinical Trial}.
\newblock {\em JAMA Network Open}, 2020.

\bibitem{ranoux2002respective}
D~Ranoux, C~Gury, J~Fondarai, JL~Mas, and M~Zuber.
\newblock Respective potencies of botox and dysport: a double blind, randomised, crossover study in cervical dystonia.
\newblock {\em Journal of Neurology, Neurosurgery \& Psychiatry}, 72(4):459--462, 2002.

\bibitem{lu2023deep}
Kosaku Sato and Jialu Wang.
\newblock Deep learning based multi-label image classification of protest activities.
\newblock {\em arXiv preprint arXiv:2301.04212}, 2023.

\bibitem{shen2023genocraft}
Minjie Shen, Yue Zhao, Chenhao Li, Fan Meng, Xiao Wang, David Herrington, Yue Wang, and Capucine Van~Rechem.
\newblock Genocraft: A comprehensive, user-friendly web-based platform for high-throughput omics data analysis and visualization.
\newblock {\em arXiv preprint arXiv:2312.14249}, 2023.

\bibitem{siah2021predicting}
Kien~Wei Siah, Nicholas Kelley, Steffen Ballerstedt, Bj{\"o}rn Holzhauer, Tianmeng Lyu, David Mettler, Sophie Sun, Simon Wandel, Yang Zhong, Bin Zhou, et~al.
\newblock Predicting drug approvals: The novartis data science and artificial intelligence challenge.
\newblock {\em Available at SSRN 3796530}, 2021.

\bibitem{solomon2019angiotensin}
Scott~D Solomon, John~JV McMurray, Inder~S Anand, Junbo Ge, Carolyn~SP Lam, Aldo~P Maggioni, Felipe Martinez, Milton Packer, Marc~A Pfeffer, Burkert Pieske, et~al.
\newblock Angiotensin--neprilysin inhibition in heart failure with preserved ejection fraction.
\newblock {\em New England Journal of Medicine}, 381(17):1609--1620, 2019.

\bibitem{solomon2017angiotensin}
Scott~D Solomon, Adel~R Rizkala, Jianjian Gong, Wenyan Wang, Inder~S Anand, Junbo Ge, Carolyn~SP Lam, Aldo~P Maggioni, Felipe Martinez, Milton Packer, et~al.
\newblock Angiotensin receptor neprilysin inhibition in heart failure with preserved ejection fraction: rationale and design of the paragon-hf trial.
\newblock {\em JACC: Heart Failure}, 5(7):471--482, 2017.

\bibitem{wang2024reflection}
Hanyin Wang, Chufan Gao, and Jimeng Sun.
\newblock A reflection and outlook on clinical adaption of large language models.
\newblock In {\em AAAI 2024 Spring Symposium on Clinical Foundation Models}, 2024.

\bibitem{wang2024twin}
Yue Wang, Yinlong Xu, Zihan Ma, Hongxia Xu, Bang Du, Honghao Gao, and Jian Wu.
\newblock {TWIN-GPT}: Digital twins for clinical trials via large language model.
\newblock {\em arXiv preprint arXiv:2404.01273}, 2024.

\bibitem{wang2023meditab}
Zifeng Wang, Chufan Gao, Cao Xiao, and Jimeng Sun.
\newblock Meditab: Scaling medical tabular data predictors via data consolidation, enrichment, and refinement, 2023.

\bibitem{wang2022transtab}
Zifeng Wang and Jimeng Sun.
\newblock Transtab: Learning transferable tabular transformers across tables.
\newblock {\em Advances in Neural Information Processing Systems}, 35:2902--2915, 2022.

\bibitem{wishart2018drugbank}
David~S Wishart et~al.
\newblock Drugbank 5.0: a major update to the drugbank database for 2018.
\newblock {\em Nucleic acids research}, 2018.

\bibitem{wolf-etal-2020-transformers}
Thomas Wolf, Lysandre Debut, Victor Sanh, Julien Chaumond, Clement Delangue, Anthony Moi, Pierric Cistac, Tim Rault, Rémi Louf, Morgan Funtowicz, Joe Davison, Sam Shleifer, Patrick von Platen, Clara Ma, Yacine Jernite, Julien Plu, Canwen Xu, Teven~Le Scao, Sylvain Gugger, Mariama Drame, Quentin Lhoest, and Alexander~M. Rush.
\newblock Transformers: State-of-the-art natural language processing.
\newblock In {\em Proceedings of the 2020 Conference on Empirical Methods in Natural Language Processing: System Demonstrations}, pages 38--45, Online, October 2020. Association for Computational Linguistics.

\bibitem{wong2019estimation}
Chi~Heem Wong, Kien~Wei Siah, and Andrew~W Lo.
\newblock Estimation of clinical trial success rates and related parameters.
\newblock {\em Biostatistics}, 20(2):273--286, 2019.

\bibitem{lu2022cot}
Chiung-Ting Wu, Sarah~J Parker, Zuolin Cheng, Georgia Saylor, Jennifer~E Van~Eyk, Guoqiang Yu, Robert Clarke, David~M Herrington, and Yue Wang.
\newblock Cot: an efficient and accurate method for detecting marker genes among many subtypes.
\newblock {\em Bioinformatics Advances}, 2(1):vbac037, 2022.

\bibitem{wu2012identifying}
Yonghui Wu et~al.
\newblock Identifying the status of genetic lesions in cancer clinical trial documents using machine learning.
\newblock {\em BMC genomics}, 2012.

\bibitem{ye2009stochastic}
Jerry Ye, Jyh-Herng Chow, Jiang Chen, and Zhaohui Zheng.
\newblock Stochastic gradient boosted distributed decision trees.
\newblock In {\em CIKM}, 2009.

\bibitem{yi2018enhance}
Steven Yi, Minta Lu, Adam Yee, John Harmon, Frank Meng, and Saurabh Hinduja.
\newblock Enhance wound healing monitoring through a thermal imaging based smartphone app.
\newblock In {\em Medical Imaging 2018: Imaging Informatics for Healthcare, Research, and Applications}, volume 10579, pages 438--441. SPIE, 2018.

\bibitem{zhang2020deepenroll}
Xingyao Zhang et~al.
\newblock Patient-trial matching with deep embedding and entailment prediction.
\newblock In {\em WWW}, 2020.

\end{thebibliography}

\end{document}